\documentclass[useAMS,usegraphicx,usenatbib,usedcolumn]{mn2e}
\usepackage{amsmath}
\usepackage{amssymb}

\title[Demonstration of polaroastrometry]{On-sky demonstration of optical polaroastrometry}

\author[B.~Safonov]{B.~Safonov\thanks{E-mail: safonov@sai.msu.ru}\\
Lomonosov Moscow State University,Sternberg Astronomical Institute, Universitetsky prosp. 13, 119992 Moscow, Russia}

\begin{document}
\date{Accepted --- Received ---}
\pagerange{\pageref{firstpage}--\pageref{lastpage}}
\pubyear{2015}
\maketitle
\label{firstpage}

\begin{abstract}
A method for measuring the difference between centroids of polarized flux and total flux of an astronomical object -- {\it polaroastrometry} -- is proposed. The deviation of the centroid of flux corresponding to Stokes parameter $Q$ or $U$ from the centroid of total flux multiplied by dimensionless Stokes parameter $q$ or $u$ respectively, was used as a signal. The efficiency of the method is demonstrated on the basis of observations made in the $V$ band by using an instrument combining features of a two-beam polarimeter with a rotating half-wave plate and a speckle interferometer. The polaroastrometric signal noise is 60--70~$\mu$as rms for a total number of accumulated photoelectrons $N_e$ of $10^9$ from a 70-cm telescope; this corresponds to a total integration time of 500~sec and an object magnitude $V=6$~mag. At smaller $N_e$ the noise increases as $\approx 1.7^{\prime\prime}/\sqrt{N_e}$, while at larger $N_e$ it remains the same owing to imperfection of the half-wave plate. For main sequence stars that are unpolarized and polarized by interstellar dust and the Mira type variable R~Tri the signal was undetectable. For the Mira type variable $\chi$~Cyg the polaroastrometric signal is found to be $310\pm70$ and $300\pm70$~$\mu$as for Stokes $Q$ and $U$ respectively; for $o$~Cet these values are $490\pm100$ and $1160\pm100$~$\mu$as. The significant value of the polaroastrometric signal provides evidence of the asymmetry of the polarized flux distribution.
\end{abstract}
 
\begin{keywords}
instrumentation: polarimeters -- techniques: polarimetric -- stars: AGB and post-AGB
\end{keywords}

\section{Introduction}

Quite frequently, some fraction of the radiation of an astronomical object becomes polarized as a result of physical conditions and/or its intrinsic asymmetry and can be measured by means of conventional polarimetry \citep{TinbergenBook}. For the same reasons the centroid of an object can depend on the orientation of the transmitted polarization. 

This effect is akin to the dependence of the centroid on the wavelength of observation, the measurement of which constitutes the subject of spectroastrometry, as proposed by \citet{Beckers1982}. The spectroastrometric signal can be measured on scales smaller than the diffraction limit of the telescope and its analysis allows one to impose additional constraints on the models of binary stars \citep{Bailey1998}, circumstellar discs \citep{Pontoppidan2011}, and active galactic nuclei \citep{Gnerucci2013}. 

Similarly, measurement of the deviation of the centroid of the polarized flux from the centroid of the total flux -- polarimetric astrometry, which we call for brevity {\it polaroastrometry} -- may provide essentially new information about the object \citep{Johnson2014,Safonov2013}.

% of polarized flux can be displaced with respect to the centroid of total flux just as centroid can depend on wavelength of radiation.

% in this case mentioned displacement delivers essentially new information about the object.

It is convenient to characterize partially polarized light by the so called Stokes parameters $I, Q, U$, and $V$, which are related to the polarization ellipse in the following way:
\begin{align}\begin{split}
I & = I, \\
Q & = I p \cos2\chi \cos2\xi, \\
U & = I p \sin2\chi \cos2\xi, \\
V & = I p \sin2\xi, 
\label{eq:stokesDef}
\end{split}\end{align}
where $I$ is the total intensity, $p$ is the fraction of polarized light, $\chi$ is the angle of polarization, and $\xi$ is the angle characterizing elliptical polarization. $\xi$ varies from $-\pi/4$ to $\pi/4$, for linearly polarized light $\xi=0$, for circularly polarized light $\xi=\pm\pi/4$. Eqs. (\ref{eq:stokesDef}) show that the parameters $Q$ and $U$ define linear polarization and $V$ defines circular one. The Stokes parameters are preferable because they have the same units of measurement as the intensity.

The appearance of an astronomical object can be specified as a dependence of its intensity on the angular coordinate on the sky $O_I(\boldsymbol{\alpha})$; by analogy it is possible to introduce the dependencies of other Stokes parameters on $\boldsymbol{\alpha}$: $O_Q(\boldsymbol{\alpha})$, $O_U(\boldsymbol{\alpha})$, and $O_V(\boldsymbol{\alpha})$.

The total Stokes parameters $Q,\,U$, and $V$ are integrals of the functions $O_Q(\boldsymbol{\alpha})$, $O_U(\boldsymbol{\alpha})$, and $O_V(\boldsymbol{\alpha})$ over $\boldsymbol{\alpha}$ and are measured by means of polarimetry. Usually dimensionless Stokes parameters are considered: $q=Q/I$, $u=U/I$, and $v=V/I$.

Consider the images of an object obtained by an ideal two-beam polarimeter that splits light into horizontally and vertically polarized components:
\begin{equation}
F_{\mathrm{h}}(\boldsymbol{\alpha}) = O_I(\boldsymbol{\alpha}) + O_Q(\boldsymbol{\alpha}),
\label{eq:simpleTwobeam1}
\end{equation}
\begin{equation}
F_{\mathrm{v}}(\boldsymbol{\alpha}) = O_I(\boldsymbol{\alpha}) - O_Q(\boldsymbol{\alpha}).
\label{eq:simpleTwobeam2}
\end{equation}

The centroids of these images are:
\begin{equation}
\boldsymbol{\alpha}_{\mathrm{h}} = \frac{\int \boldsymbol{\alpha} \bigl(O_{I}(\boldsymbol{\alpha}) + O_{Q}(\boldsymbol{\alpha})\bigr) d\boldsymbol{\alpha}}{\int \bigl(O_{I}(\boldsymbol{\alpha}) + O_{Q}(\boldsymbol{\alpha})\bigr) d\boldsymbol{\alpha}},
\end{equation}
\begin{equation}
\boldsymbol{\alpha}_{\mathrm{v}} = \frac{\int \boldsymbol{\alpha} \bigl(O_{I}(\boldsymbol{\alpha}) - O_{Q}(\boldsymbol{\alpha})\bigr) d\boldsymbol{\alpha}}{\int \bigl(O_{I}(\boldsymbol{\alpha}) - O_{Q}(\boldsymbol{\alpha})\bigr) d\boldsymbol{\alpha}}.
\end{equation}
The half--difference of these centroids $\boldsymbol\Delta_Q = (\boldsymbol{\alpha}_{\mathrm{h}} - \boldsymbol{\alpha}_{\mathrm{v}})/2$ can be measured with significant accuracy, as will be proven below. The half-difference is connected to the polarization properties of the object in a simple way (if one assumes $q \ll 1$):
\begin{equation}
\boldsymbol\Delta_Q = q \Biggl[ \frac{\int \boldsymbol{\alpha} O_{Q}(\boldsymbol{\alpha}) d\boldsymbol{\alpha}}{\int O_{Q}(\boldsymbol{\alpha}) d\boldsymbol{\alpha}} - \frac{\int \boldsymbol{\alpha} O_{I}(\boldsymbol{\alpha}) d\boldsymbol{\alpha}}{\int O_{I}(\boldsymbol{\alpha}) d\boldsymbol{\alpha}} \Biggr].
\end{equation}
Similar values can be defined for the other two Stokes parameters $U$ and $V$. The variables $\boldsymbol{\Delta}_Q$, $\boldsymbol{\Delta}_U$, and $\boldsymbol{\Delta}_V$ that constitute the {\it polaroastrometric signal} are the basic observables of polaroastrometry. Let us denote the components of these vectors as follows: $\boldsymbol{\Delta}_Q = (s_q^{\star},t_q^{\star})$, $\boldsymbol{\Delta}_U = (s_u^{\star},t_u^{\star})$, and $\boldsymbol{\Delta}_V = (s_v^{\star},t_v^{\star})$. 

By knowing the dimensionless Stokes parameters for our object, and given that they significantly deviate from zero, it is possible to compute the deviations of the centroids of polarized flux from the centroid of the total flux, {\it polaricentres} $\boldsymbol{\alpha}_Q,\boldsymbol{\alpha}_U$, and $\boldsymbol{\alpha}_V$, from the polaroastrometric signal:
\begin{equation}
\boldsymbol{\alpha}_Q = \boldsymbol{\Delta}_Q / q,\,\,\,\, \boldsymbol{\alpha}_U = \boldsymbol{\Delta}_U / u,\,\,\,\, \boldsymbol{\alpha}_V = \boldsymbol{\Delta}_V / v.
\end{equation}
The components of polaricentres will be denoted: $\boldsymbol{\alpha}_Q = (s_q,t_q)$, $\boldsymbol{\alpha}_U = (s_u,t_u)$, and $\boldsymbol{\alpha}_V = (s_v,t_v)$. This derivation is correct for Stokes parameters $Q, U$, and $V$, but in following we will consider only $Q$ and $U$, which are related to linear polarization.

%(here three equations are given simultaneously, we iterate through the Stokes parameters in subscript)
%\begin{equation}
%\boldsymbol{\alpha}_{(Q,U,V)} = \frac{\int \boldsymbol{\alpha} O_{(Q,U,V)}(\boldsymbol{\alpha}) d\boldsymbol{\alpha}}{\int O_{(Q,U,V)}%(\boldsymbol{\alpha}) d\boldsymbol{\alpha}}.
%\label{eq:centerIdef}
%\end{equation}

\begin{figure}
	\begin{center}
	\includegraphics[width=8cm]{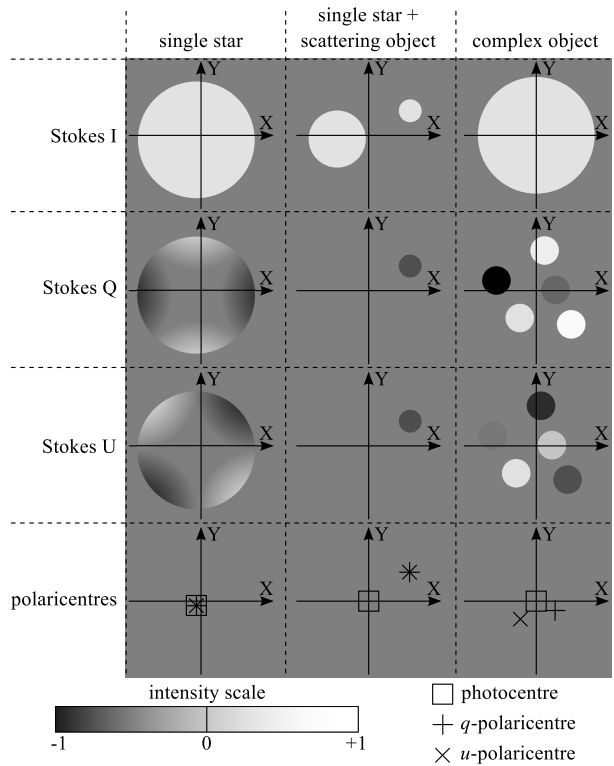}
	\end{center}
	\caption{
	Different types of objects from the point of view of polaroastrometry. Shown in rows are Stokes parameters $I$, $Q$, $U$, photocentres (squares) and polaricentres (crosses). Columns display the cases of a single star, a single star with a scattering object nearby, and a complex object.
	\label{fig:explanation}}
\end{figure}

Let us consider examples of astrophysical objects that differ from the observational point of view for the polaroastrometry.
\begin{enumerate}
\item If the distribution of the polarized flux of the object is centrally symmetrical (e.g., a main sequence star), or there is no polarized light, then the polaricentres coincide with the photocentre and the polaroastrometric signal equals zero (first column in Fig. \ref{fig:explanation}). 
\item If the object is polarized by interstellar dust, then the polaricentres coincide with the photocentre\footnote{Strictly speaking, this is correct only in absence of the atmospheric dispersion, see Section \ref{subs:atmdisp}} (first column in Fig. \ref{fig:explanation}).
\item If the source of polarized light is confined within a small region (relative to the precision of the method), then the polaricentres will coincide with each other, but deviate from the photocentre (second column in Fig. \ref{fig:explanation}).
\item In other cases the coincidence of polaricentres is possible only for special and therefore unlikely object geometry (third column in Fig. \ref{fig:explanation}). For complex objects the polaricentres can deviate from the photocentre even when the total polarization is close to zero.
\end{enumerate}
We can conclude that even a qualitative analysis of polaroastrometric data -- the detection of a deviation of polaricentres from the photocentre or coincidence of polaricentres -- presents a valuable tool for astronomical object diagnosis. A quantitative interpretation of the polaroastrometric signal is also possible by using a priori knowledge about an object.

\begin{figure*}
	\centering
	\includegraphics[width=14cm]{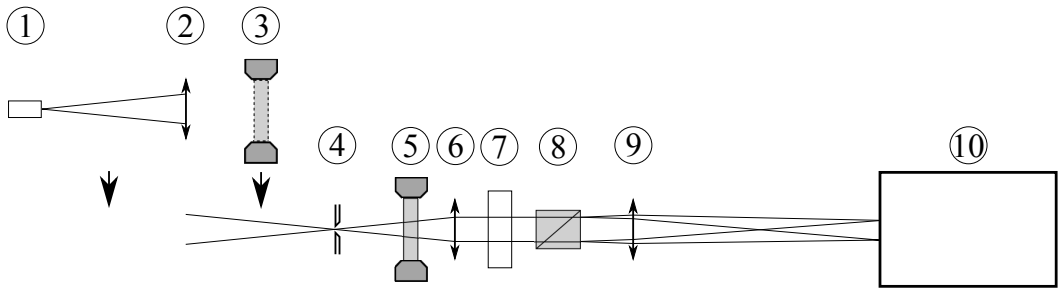}
	\caption{
	Schematic of the instrument used in the experiment. Numbers are for the following components: (1) pinhole, (2) auxiliary lens, forming an image of the pinhole at the first focal plane (with components 1 and 2 being assembled as a single unit which that be inserted into the beam manually), (3) linear polarizer, which is inserted into the beam manually, (4) field diaphragm, (5) half-wave plate, (6) collimator lens, (7) filter, (8) Wollaston prism, (9) camera lens, and (10) detector.
	\label{fig:setupTel}}
\end{figure*}

Here we propose a method of optical polaroastrometry and study its performance on the basis of observations made with the prototype of the Multimode Fast Camera installed on the 70-cm telescope in Moscow (see description in Section \ref{sec:instrument}). The method used to estimate the dimensionless Stokes parameters $q$ and $u$ from the observational data is presented in Section \ref{sec:polarimetry}. The polaroastrometric signal is extracted from the same data by using differential speckle polarimetry \citep{Safonov2013}. Details of the method in an application to polaroastrometry and a noise analysis are given in Section \ref{sec:polaroastrometry}. The results in the form of dimensionless Stokes parameters and the polaroastrometric signal are given in Section \ref{sec:results} for unpolarized stars, stars polarized by interstellar dust and stars presumably possessing a detectable signal. Section \ref{sec:conclusion} presents conclusions. Some auxiliary results are presented in Appendices \ref{app:hwp} and \ref{app:rotation}.

\section{Instrument}
\label{sec:instrument}

For the polaroastrometric experiment we used a prototype of the Multimode Fast Camera under design for the 2.5-m telescope of the Sternberg Astronomical Institute (SAI). A schematic of the instrument is depicted in Fig. \ref{fig:setupTel}. The instrument combines features of a two-beam polarimeter with a rotating half-wave plate and a speckle interferometer.

We used an Andor iXon 897 electron multiplying CCD (EMCCD) camera as the detector. Owing to electron multiplication technology, readout noise in this type of detector is greatly reduced, at the cost of a twofold increase in photon noise (so--called multiplication noise). The frame rate was chosen to be 35 frames per second, which is fast enough to reduce blurring of the image by atmospheric turbulence.

The first optical component of the system is a half-wave plate (HWP), which effectively rotates the polarization plane of incoming radiation. This has two goals: measurement of both parameters of linear polarization and calibration of the instrumental polarization of the optics located after the HWP. The HWP (Edmund Optics, EO-49232) consists of several polymer layers selected so that the total phase retardance between ordinary and extraordinary beams deviates from $\pi$~rad by $\leq 0.02$~rad inside the $V$ band. A motorized rotation stage Standa~8MRU was used to control the position angle of the HWP.

Two achromatic lenses with a collimated beam between them were used as relay optics. These lenses give a magnification of $\approx4.6$. A $V$ filter and a Wollaston prism (RIVoptics) were installed in the collimated beam. The prism splits the beam in two, polarizing it horizontally and vertically.
 
\begin{figure}
\begin{center}
\includegraphics[width=8cm]{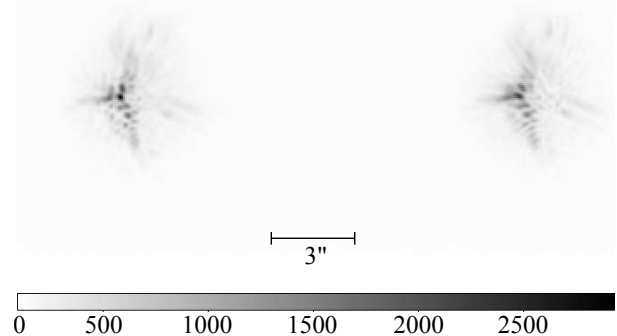}
\end{center}
\caption{
A single frame from Vega, obtained with the prototype of the Multimode Fast Camera and an exposure 30~ms. The separation of the images is 210~pixels. The brightness scale at the bottom below is in counts.
\label{fig:vega_single}
}
\end{figure}

A slit diaphragm is installed in the first focal plane; this precludes any overlap of the beams split by the Wollaston prism and limits the field of view to $13^{\prime\prime}\times33^{\prime\prime}$. An example of an image obtained with the instrument is shown in Fig. \ref{fig:vega_single}. The instrument is controlled from a custom command line C++ program running under OS GNU/Linux.

The instrument was installed at the Cassegrain focus of the 70-cm telescope AZT-2 located near the SAI building in Moscow. The equivalent telescope focal length is 10.5~m; with relay optics of the instrument this is converted to $\approx48.3$~m. The angular sampling frequency is $f_d = 3.22\times10^6~\mathrm{rad}^{-1}$, whereas the cutoff frequency of the optical system is $f_c = D/\lambda = 1.27\times10^6~\mathrm{rad}^{-1}$. Therefore the Nyquist criterion $f_d>2f_c$ is fulfilled for this setup, which is necessary for correct computation of the Fourier spectrum of images.

% It was possible to rotate the instrument around the optical axis arbitrarily.  A mount type of the telescope is equatorial.

We adopted the following observation sequence. The detector continuously obtained a series of frames, without gaps between them. Meanwhile, the HWP rotated at a constant angular speed. The frame rate and a HWP speed were matched so that the HWP position angle $\theta$ changed by $9^{\circ}$ over one exposure. Because it was not possible to measure and record $\theta$ for each frame with this setup, we determined it a posteriori. For each series we introduced a linear polarizer at a known orientation into the beam for 10--15 per cent of the total series duration at its beginning and end. During these periods the polarization of incoming radiation can be considered as known and it is possible to determine corresponding $\theta$. Then $\theta$ can be interpolated in the middle of the series, when the linear polarizer was removed from the beam. Data obtained using the described technique allow the extraction of both polaroastrometry and polarimetry information.

Observations have been conducted in August--October 2014 in five sessions, with the instrument being removed from the telescope between sessions. Throughout a session the angular scale $w$ and the position angle $\phi$ of the camera were fixed by the equatorial mount of the telescope. We measured these parameters for each session for the transformation of measurements from the instrumental reference system to the equatorial J2000. Because our experiment is not very demanding in terms of $w$ and $\phi$ accuracy, for their estimation we used binary and double stars with well-known ephemerides: HIP94336, HIP103669, and HIP108917 \citep{Mason2007,Farrington2014forMNRAS}. We acquired long-exposure images of these stars and then measured a separation vector of the stellar images by approximating one image with the other. By comparing the parameters of these vectors with the ephemerides we obtained $w$ and $\phi$. The results of our measurements are given in Table \ref{table:plate}. For each session we also observed at least one polarization standard and an unpolarized star.

\begin{table}
\caption{
Angular scale $w$ and the position angle $\phi$ of the camera for the five observational sessions, where the position angle is being measured eastward from north to the readout register of the CCD camera. Dates of the session and object names used for the determination of camera parameters are given in the second and third column, respectively.
\label{table:plate}}
\centering
\begin{tabular}{ccccc}
\hline
Session & & & & \\
number & Date & Object       & $w$,~mas/pix & $\phi, ^{\circ}$ \\
\hline
1            & 3.08 --     & HIP94336  & $68.8\pm0.8$ & $69.5\pm0.3$ \\
             & 4.08        & HIP108917 & & \\
2            & 22.08 --    & HIP103669 & $68.6\pm1.7$ & $34.0\pm0.6$ \\
             & 28.08       & HIP108917 & & \\
3            & 11.09 --    & HIP94336  & $68.9\pm0.8$ & $51.6\pm0.4$ \\
             & 18.09       & HIP108917 & & \\
4            & 19.09       & HIP94336  & $69.3\pm0.8$ & $322.1\pm0.3$ \\
             &             & HIP108917 & & \\
5            & 02.10       & HIP94336  & $70.4\pm0.7$ & $51.2\pm0.3$ \\
             &             & HIP108917 & & \\
\hline
\end{tabular}
\end{table}

\section{Polarimetry}
\label{sec:polarimetry}

The instrument we used is a two-beam polarimeter, whose main advantage is that it is insensitive to variations in atmospheric transparency and to stellar scintillation. However, it is susceptible to the difference in transmission between the two beams of the polarimeter. To mitigate this effect, switching of the images corresponding to orthogonal polarizations is employed. The switching usually is performed with the help of an HWP installed before the polarimeter \citep{Bagnulo2009}. In our case, images are switched every 140~ms.

One potential drawback of this scheme is its high sensitivity to any imperfection of the HWP. Nevertheless, it is possible to estimate this effect by using a generalization of the double difference method as described in the following. This generalization is closely related to the adopted methodology of polaroastrometry (Section \ref{sec:polaroastrometry}). 

Preparatory processing of data is performed as follows. First, we average frames with $\theta$ differing by $\leq 3^{\circ}$, which is small compared to the angle through which the HWP passes during one exposure. Then we subtract a bias and a constant background from the averaged frames. Fluxes corresponding to two images, $J_{Lk}$ and $J_{Rk}$, are extracted by using aperture photometry, where $k$ is the number of the averaged image. The relative positions of the apertures for photometry are the same for all frames.

In Appendix \ref{app:hwp} the fluxes are shown to be connected to the Stokes parameters of the object in the instrument reference system $I, Q_d, U_d$, and $V_d$ as follows:
\begin{multline}
J_{Lk} = W(1+\Delta W_k/2) K_k  \\ \times \biggl[I + \zeta \bigl[ Q_d \cos(4(\theta_k+\kappa(\theta_k))) + U_d  \sin(4(\theta_k+\kappa(\theta_k))) \bigr] \\ + \zeta^{\prime} V_d \delta(\theta_k) \sin(2(\theta_k+\kappa(\theta_k)))\biggr] + \nu_{Lk},
\label{eq:leftFlux}
\end{multline}
\begin{multline}
J_{Rk} = W(1-\Delta W_k/2) K_k \\ \times \biggl[I - \zeta \bigl[ Q_d \cos(4(\theta_k+\kappa(\theta_k))) + U_d  \sin(4(\theta_k+\kappa(\theta_k))) \bigr] \\ - \zeta^{\prime} V_d \delta(\theta_k) \sin(2(\theta_k+\kappa(\theta_k)))\biggr] + \nu_{Rk}.
\label{eq:rightFlux}
\end{multline}
Here $\theta_k$ is the mean angle $\theta$ over frame $k$ duration, $K_k$ is the mean atmospheric transparency, $W$ is the total transmission of the optical system, $\Delta W_k$ is the difference in transmission for the polarimeter beams, which includes a flat field error. The values $\delta(\theta_k)$ and $\kappa(\theta_k)$ are the parameters of the HWP imperfection; their absolute values are relatively  small (see appendix \ref{app:hwp} for details) $\nu_{Lk}$ and $\nu_{Rk}$ characterize noise that is independent for separate frames, e.g., photon noise. % DO NOT DELETE THIS Please note that some functions (e.g. $\Delta W, K$) depend on $k$ explicitly, while the other through the dependence on $\theta$. This is done intentionally to emphasize that functions of second type have the same values for $\theta_{k}=\theta_{k}$ and $k \ne l$. For the functions of first type this is not true, nevertheless they can be considered as functions of $\theta$, but {\it random} functions.

Coefficients $\zeta$ and  $\zeta^{\prime}$ are present in Eq. (\ref{eq:leftFlux}) and (\ref{eq:rightFlux}) because $\theta$ as well as its cosine and sine, changes during the course of obtaining the frame, and therefore some averaging occurs. $\zeta$ gives the sense of the polarimeter {\it effectiveness} \citep{TinbergenBook} and equals $\zeta\approx1-(2/3)\theta_f^2$, where $\theta_f$ is the angle through which the HWP passes during the exposure. At $\theta_f=9^{\circ}$ the coefficient $\zeta$ equals $0.984$.

From $J_{Lk}$ and $J_{Rk}$ the $P_k$ values can be computed as follows:
\begin{equation}
P_k =  \frac{J_{Lk} - J_{Rk}}{J_{Lk} + J_{Rk}}.
\end{equation}
Substitution of expressions for the fluxes (\ref{eq:leftFlux}) and (\ref{eq:rightFlux}) gives
\begin{multline}
P(\theta) = \zeta (q_d \cos(4\theta) + u_d \sin(4\theta)) \\ + \sum_{n=0}^{\infty} \omega_q(n) \cos(n\theta) + \sum_{n=0}^{\infty} \omega_u(n) \sin(n\theta).
\label{eq:polfrac}
\end{multline}
Here the functions $\Delta W, \delta$, and $\kappa$ describing instrumental effects are decomposed into a Fourier series over $\theta$ and grouped into the last two sums. In the following we designate the Fourier spectrum of $P(\theta)$ as $\widetilde{P}(f_{\theta})$.

From Eq. (\ref{eq:polfrac}) one can see the difference between a signal carrying information about the linear polarization of the source and instrumental effects such as variations of the system transmission and the HWP imperfection. The signal occupies only one frequency in the spectrum $\widetilde{P}(f_{\theta})$; meanwhile, the noise is distributed over all frequencies, including the signal frequency. Therefore, the linear polarization can be estimated as the amplitude of fourth harmonics of $\widetilde{P}(f_{\theta})$, but it will be inevitably biased by unknown values $\omega_{u,q}(4)$. All that remains is to estimate their amplitudes by interpolation and take these as a measure of the uncertainty of the $q_d$ and $u_d$ estimation.

\begin{figure}
	\centering
	\includegraphics[width=8cm]{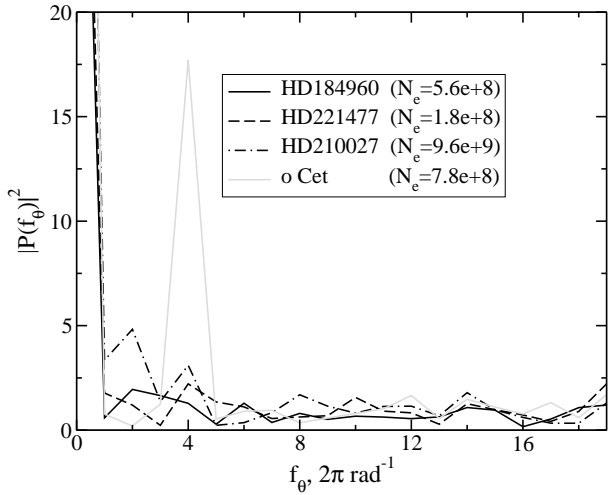}
	\caption{
	Spectrum modulus squared, $|\widetilde{P}(f_{\theta})|^{2}$ normalized by the expected level of the photon noise for four objects. Black lines are for presumably unpolarized stars; grey one is for weakly polarized $o$~Cet. The legend gives the total numbers of accumulated photoelectrons.
	\label{fig:totalpolSpec}
	}
\end{figure}

In Fig. \ref{fig:totalpolSpec} the spectra $|\widetilde{P}(f_{\theta})|^{2}$ for three presumably unpolarized stars are given. As unpolarized stars we have taken main sequence stars at distances $<66$~pc and galactic latitude $>10^{\circ}$, as was done by \citet{Berdyugina2011}. One can see that the spectra are quite uniform, and that there are no marked frequencies, which demonstrates the possibility of estimation of the fourth harmonic uncertainty. It is notable that $|\widetilde{P}(f_{\theta})|^{2}$ slowly rises at low frequencies for the bright star. This effect can be explained by the fact that the photon component of the noise for this star is sufficiently low to make the instrumental effects visible. 

%DO NOT DELETE THIS A detailed theoretical and experimental analysis of noise of $\widetilde{P}(f_{\theta})$ is of great interest but falls beyond the scope of this work. It will be performed in an application to polaroastrometry, Subsection \ref{subs:noise}.

The spectrum $|\widetilde{P}(f_{\theta})|^{2}$ for $o$~Cet (2014 October 2) is also presented in Fig. \ref{fig:totalpolSpec}. In this case one sees that the fourth harmonic signal is significantly higher than the noise, indicating the presence of detectable linear polarization. For polarization standard stars the fourth harmonic amplitude exceeds the noise level by a factor of 100.

The advantage of this method of spectral decomposition of $P(\theta)$ over the double difference method \citep{Bagnulo2009} is that the former enables an adequate estimation of the uncertainty of the resulting value caused by instrumental effects, especially the HWP imperfection (see Appendix \ref{app:hwp}). For polarimetry such an estimation is important only for the brightest stars, but for polaroastrometry the sensitivity to the HWP imperfection is significantly stronger and a more accurate estimation of its effect becomes critical. A similar approach to analysing data from polarimeters with rotating HWPs was applied in millimeter waves by \citep{Johnson2007forMNRAS}.

From $q_d$ and $u_d$ one can estimate the degree of polarization $p_d$ and the angle of polarization $\chi_d$; the latter can then be transformed to the equatorial coordinate system.

To verify the method and test its stability we observed polarization standards and presumably unpolarized stars in every session. The results are given in Table \ref{table:polst} in the form of $p$ and $\chi$. For comparison, values obtained by \citet{Schmidt1992} are given. For the standards, the degree of polarization coincides with a precision 3 per cent, and the angle of polarization with a precision $1\div2^{\circ}$, both of which are acceptable. The stability of $p$ is 1 per cent, and that of $\chi$ is $1^{\circ}$; the difference between sessions exceeds the errors insignificantly.
 
The small value of the measured polarization for the unpolarized stars provides evidence that the instrumental polarization of the telescope is small and can be neglected, which is expected because the optical design is centrally symmetrical.

The polarimetry results are also given in the form of dimensionless Stokes parameters $q$ and $u$ in Table \ref{table:polaroastrometry}.

\begin{table}
\caption{
Comparison of measured and expected polarization parameters of standards. The first column lists the number in Henry Draper catalogue or Bayer name; the second lists S for observational session number and the E for value from \citet{Schmidt1992}; the third lists degree of polarization $p$ and the fourth lists the polarization angle $\chi$. For stars with small polarization, reliable estimation of $\chi$ cannot be made.
\label{table:polst}
}
\centering
\begin{tabular}{cccc}
\hline
HD/name     & S/E      & $p\times10^{4}$ & $\chi, ^{\circ}$  \\
\hline

HD7927 	& S1 & $332.9\pm0.3$ & $93.1\pm0.3$ \\
$\phi$ Cas	& S3 & $334.5\pm0.5$ & $92.7\pm0.4$ \\
	& S4 & $330.3\pm0.8$ & $93.1\pm0.3$ \\
        & E  & $329.8\pm2.5$ & $91.1\pm0.2$  \\
HD204827 	& S1 & $547.7\pm1.8$ & $59.1\pm0.3$ \\ % astrom4
 		& S2 & $548.8\pm2.8$ & $60.5\pm0.7$ \\ % plate1
	        & S3 & $547.7\pm2.7$ & $59.9\pm0.4$ \\ % astrom9
		& S4 & $549.7\pm1.7$ & $60.5\pm0.3$ \\ % astrom11
		& S5 & $554.4\pm1.7$ & $61.5\pm0.3$ \\ % astrom12
                & E  & $532.2\pm1.4$ & $58.7\pm0.1$ \\
HD184960 	& S1 & $1.8\pm0.8$ & \\ % HIP96258
HD221477	& S1 & $3.5\pm3.3$ & \\ % HIP116174
HD210027 	& S5 & $1.3\pm0.5$ & \\	% HIP109176
\hline
\end{tabular}
\end{table}

\section{Polaroastrometry}
\label{sec:polaroastrometry}

\subsection{Basics}
\label{subs:pa:basics}

To take into account the effect of the HWP the model of images in the two-beam polarimeter (\ref{eq:simpleTwobeam1}) and (\ref{eq:simpleTwobeam2}) should be extended as follows:
\begin{equation}
F_L(\boldsymbol{\alpha}) = O_I(\boldsymbol{\alpha}) + a(\theta) O_Q(\boldsymbol{\alpha}) + b(\theta) O_U(\boldsymbol{\alpha}),
\end{equation}
\begin{equation}
F_R(\boldsymbol{\alpha}) = O_I(\boldsymbol{\alpha}) - a(\theta) O_Q(\boldsymbol{\alpha}) - b(\theta) O_U(\boldsymbol{\alpha}).
\end{equation}
In our case, $a(\theta) = \zeta \cos(4\theta)$ and $b(\theta) = \zeta \sin(4\theta)$, where $\theta$ is the orientation of the HWP. The polaroastrometric signal is
\begin{multline}
\boldsymbol{\Delta}(\theta) = \frac{1}{2}\Biggl[\frac{\boldsymbol{\alpha}_I+a(\theta)\boldsymbol{\alpha}_Q+b(\theta)\boldsymbol{\alpha}_U}{I+a(\theta)Q+b(\theta)U} \\ - \frac{\boldsymbol{\alpha}_I-a(\theta)\boldsymbol{\alpha}_Q-b(\theta)\boldsymbol{\alpha}_U}{I-a(\theta)Q-b(\theta)U}\Biggr].
\end{multline}
By supposing that $q^2+u^2\ll1$ (which is correct for most astronomical objects), it is possible to derive
\begin{equation}
\boldsymbol{\Delta}(\theta) = a(\theta) \boldsymbol{\Delta}_Q + b(\theta) \boldsymbol{\Delta}_U.
\label{eq:pasignal}
\end{equation}
In the following section we show how the values $\boldsymbol{\Delta}(\theta)$, $\boldsymbol{\Delta}_Q$ and $\boldsymbol{\Delta}_U$ can be estimated from the real data.

According to a convention accepted in optics, a beam with a Stokes vector $(I,I,0,0)$ is polarized horizontally, i.e., the plane of polarization is oriented along the OX axis. In astronomical polarimetry for a such beam the polarization plane will be oriented towards the north celestial pole (NCP). Therefore, for derivations we assume that the OX axis is directed towards the NCP. Then, for an object with $s_q > 0,t_q = 0$ the $q$-polaricentre is displaced towards north; for an object with $s_q = 0,t_q > 0$ -- towards east, which may seem counterintuitive.

\subsection{Extraction of the polaroastrometric signal}
\label{subs:pa:extraction}

The polaroastrometric signal $\boldsymbol{\Delta}$ can be extracted from the observational data in various ways, e.g. by performing a simple centroid computation or by approximating one image with another. We solve this task by using a differential speckle polarimetry (DSP), a method similar to the double difference approach \citep{Bagnulo2009}, but operating in the Fourier domain \citep{Safonov2013}.

A principal observable of the DSP method is $\mathcal{R}(\boldsymbol{f})$:
\begin{equation}
\mathcal{R}(\boldsymbol{f}) = \frac{\langle \widetilde{F}_L(\boldsymbol{f}) \widetilde{F}_R^*(\boldsymbol{f}) \rangle}{\langle \widetilde{F}_R(\boldsymbol{f}) \widetilde{F}_R^*(\boldsymbol{f}) \rangle - N_{eR}^{-1}},
\label{eq:Rl_def}
\end{equation}
where $\boldsymbol{f}$ is the spatial frequency, $\widetilde{F}_L(\boldsymbol{f})$ and $\widetilde{F}_R(\boldsymbol{f})$ are Fourier spectra of the images formed by the left and right beams of the Wollaston prism, $N_{eR}$ is the averaged number of photons forming the right image. The averaging in this equation takes place over a finite number of frames.

The computation of $\mathcal{R}(\boldsymbol{f})$ starts with removing bias, constant background and equalizing to zero pixels having counts $< 5$ RMS of read-out noise. Then two rectangular windows of $100\times100$~pixels each are selected. The left window is centred on the brightest pixel of the left image; the right window is displaced relative to the left by a constant vector. The Fourier spectra are evaluated separately for the left and right images for many frames and are combined using Eq. (\ref{eq:Rl_def}). Only images having $\theta$ differing by $\leq 3^{\circ}$ take part in the combination. In such a manner a set of $\mathcal{R}(\theta,\boldsymbol{f})$ for different $\theta$ values is formed.

For processing $\mathcal{R}(\theta,\boldsymbol{f})$ data we need a model of its formation. As a starting point we use the main result of \citep{Safonov2013}: $\mathcal{R}$ averaged over the ensemble of samples of random atmospheric phase and photon noise is
\begin{equation}
\overline{\mathcal{R}}(\boldsymbol{f}) = (1+\Delta\mathcal{R}(\boldsymbol{f})) \mathcal{R}_0(\boldsymbol{f}),
\label{eq:Rl_simple1}
\end{equation}
where $\mathcal{R}_0$ is the ratio of visibilities of the object corresponding to left and right beams, and $\Delta\mathcal{R}(\boldsymbol{f})$ is derived by the differential polarization aberrations in the system. This expression accounts for three effects: atmospheric optical turbulence, instrumental polarization and photon noise.

To adequately describe our experiment the following instrumental effects additionally need to be taken into account: (1) deviations of the beams by the Wollaston prism, (2) dispersion from the Wollaston prism and the atmosphere, (3) differential aberrations of the HWP, and (4) transformation of the polarization state by the HWP.

The first two effects are described by multiplying the Fourier spectra of images by their respective filters:
\begin{equation}
\widetilde{F}_L^{\prime}(\boldsymbol{f}) = \widetilde{F}_L(\boldsymbol{f}) D_L(\boldsymbol{f}) e^{i 2 \pi (\boldsymbol{\rho}_L\cdot\boldsymbol{f})},
\end{equation}
\begin{equation}
\widetilde{F}_R^{\prime}(\boldsymbol{f}) = \widetilde{F}_R(\boldsymbol{f}) D_R(\boldsymbol{f}) e^{i 2 \pi (\boldsymbol{\rho}_R\cdot\boldsymbol{f})},
\end{equation}
where $D_L(\boldsymbol{f})$ and $D_R(\boldsymbol{f})$ are the filters describing the dispersion, $\boldsymbol{\rho}_L$ and $\boldsymbol{\rho}_R$ account for image displacements in the left and right beams of the Wollaston prism. Neither effect depends on the angle of the HWP as long as atmospheric dispersion affects different polarization states equally and the Wollaston prism is installed after the HWP.

After substituting $\widetilde{F}_L^{\prime}(\boldsymbol{f})$ and $\widetilde{F}_R^{\prime}(\boldsymbol{f})$ into Eq. (\ref{eq:Rl_def}) relation (\ref{eq:Rl_simple1}) is modified as follows:
\begin{equation}
\overline{\mathcal{R}}(\boldsymbol{f}) = (1+\Delta\mathcal{R}(\boldsymbol{f})) \mathcal{R}_0(\boldsymbol{f}) D(\boldsymbol{f}) e^{i2\pi(\boldsymbol{\rho}\cdot\boldsymbol{f})},
\label{eq:Rl_simple2}
\end{equation}
where $D(\boldsymbol{f}) = D_L(\boldsymbol{f})/D_R(\boldsymbol{f})$ and $\boldsymbol{\rho}=\boldsymbol{\rho}_L-\boldsymbol{\rho}_R$.

By defining the telescope and instrument as two parts of an optical system separated by the HWP, it is possible to split the impact of differential aberrations into that from the telescope, the HWP, and the instrument. 
\begin{multline}
\overline{\mathcal{R}}(\theta,\boldsymbol{f}) = \mathcal{R}_0(\theta,\boldsymbol{f}) \bigl(1+\Delta\mathcal{R}_T(\theta,\boldsymbol{f}) \\ + \Delta\mathcal{R}_{\mathrm{HWP}}(\theta,\boldsymbol{f}) + \Delta\mathcal{R}_I(\boldsymbol{f})\bigr)  D(\boldsymbol{f}) e^{i2\pi(\boldsymbol{\rho}\cdot\boldsymbol{f})}.
\label{eq:Rl_full1}
\end{multline}
The correctness of this operation is provided by the additivity of bias $\Delta\mathcal{R}(\boldsymbol{f})$ \citep{Safonov2013}. Note that all the effects introduced before the HWP and inside it depend on its position angle $\theta$.

Let us suppose that some object with $\mathcal{R}_{0} \ne 1$ is confined to a region much smaller than the diffraction limit of the instrument $\lambda/D$. In this case, decomposing the exponent in the Fourier transform in a Taylor expansion and keeping only the first term, one obtains \citep{Johnson2014,Safonov2013} $\mathcal{R}_{0}(\theta) = \exp\{{i \pi (4\boldsymbol{\Delta}(\theta)\cdot\boldsymbol{f})}\}$, moreover, the exponent argument is significantly less than unity for all frequencies. Substituting $\boldsymbol{\Delta}(\theta)$ from (\ref{eq:pasignal}), we obtain:
\begin{equation}
\mathcal{R}_{0}(\theta) = \exp\bigl\{ i 4 \pi \zeta \bigl[f_x S(\theta) + f_y T(\theta) \bigr] \bigr\},
\end{equation}
where
\begin{equation}
S(\theta) = s_q^{\star} \cos(4\theta) + s_u^{\star} \sin(4\theta),
\end{equation}
\begin{equation}
T(\theta) = t_q^{\star} \cos(4\theta) + t_u^{\star} \sin(4\theta).
\end{equation}
The values $s_q^{\star}, s_u^{\star}, t_q^{\star}$, and $t_u^{\star}$ comprise the polaroastrometric signal, as defined above.

In Appendix \ref{app:hwp} it is shown that the HWP effect $\Delta\mathcal{R}_{\mathrm{HWP}}(\theta,\boldsymbol{f})$ can be represented as $1+\Delta\mathcal{R}_{\mathrm{HWP}}(\theta,\boldsymbol{f}) = \exp\{{i \phi_{\mathrm{HWP}}(\theta,\boldsymbol{f})}\}$, while $\phi_{\mathrm{HWP}} \ll 1$.

As long as the optical system of the telescope is very slow ($F/64$) and possesses central symmetry, for our level of precision $\Delta\mathcal{R}_T$ can be considered equal to unity \citep{Safonov2013}. 

By taking all these reasons into account, expression (\ref{eq:Rl_full1}) can be transformed as follows:
\begin{multline}
\overline{\mathcal{R}}(\theta,\boldsymbol{f}) = \exp\Bigl\{ i 4 \pi \zeta \bigl[f_x S(\theta) + f_y T(\theta)\bigr] + i \phi_{\mathrm{HWP}}(\theta,\boldsymbol{f}) \Bigr\} \\ \times \bigl(1+\Delta\mathcal{R}_I(\boldsymbol{f})\bigr)  D(\boldsymbol{f}) e^{i2\pi(\boldsymbol{\rho}\cdot\boldsymbol{f})}.
\label{eq:Rl_final}
\end{multline}

Eq. (\ref{eq:Rl_final}) represents the final model of the observational data. Note that real observations are not averaged over the ensemble and have some noise (which will be discussed later):
\begin{equation}
\mathcal{R}(\theta_k,\boldsymbol{f}) = \overline{\mathcal{R}}(\theta_k,\boldsymbol{f}) + \mathcal{N}_k(\boldsymbol{f}),
\end{equation}
where $\mathcal{N}_k(\boldsymbol{f})$ is the noise sample.

Model (\ref{eq:Rl_final}) allows us to formulate an algorithm for the polaroastrometric signal extraction and estimation of its error from the measurements of $\mathcal{R}(\theta,\boldsymbol{f})$. The first operation is normalizing of $\mathcal{R}(\theta,\boldsymbol{f})$ by its average over the angle $\theta$. Taking into account that the exponent argument in (\ref{eq:Rl_final}) is much less than unity we obtain:
\begin{multline}
\mathcal{R}_n(\theta,\boldsymbol{f}) = \frac{\mathcal{R}(\theta,\boldsymbol{f})}{\langle\mathcal{R}(\theta,\boldsymbol{f})\rangle_{\theta}} \\ = \exp\Bigl\{ i 4 \pi \zeta \bigl[f_x S(\theta) + f_y T(\theta)\bigr] + i \phi_{\mathrm{HWP},n}(\theta,\boldsymbol{f}) \Bigr\} \\ + \mathcal{N}_k(\boldsymbol{f}),
\label{eq:Rl_norm}
\end{multline}
where $\phi_{\mathrm{HWP},n}$ is the normalized value of $\phi_{\mathrm{HWP}}$. 

The dependence of phase $\mathcal{R}_n(\theta,\boldsymbol{f})$ on the frequency $\boldsymbol{f}$ was approximated by least squares with the plane $4\pi\bigl(c(\theta)f_x + d(\theta)f_y\bigr)$. To test validity of the model we computed a reduced chi-squared statistic, which is the sum of squared deviations of measurements from the model weighted by their variances and normalized by the number of degrees of freedom, which is the difference between the number of independent measurements of $\mathcal{R}_n(\theta,\boldsymbol{f})$ phase and number of model parameters. The reduced chi-squared turned out to be $1\div2$, which indicates the correspondence of the data to the model. 

From Eq. (\ref{eq:Rl_norm}) it follows that for the plane parameters:
\begin{equation}
c(\theta) = \zeta S(\theta) + G_n(\theta) + N_x,
\end{equation}
\begin{equation}
d(\theta) = \zeta T(\theta) + H_n(\theta) + N_y,
\end{equation}
where functions $G_n(\theta)$ and $H_n(\theta)$ characterize the HWP effect. As a result of the least squares linearity they can be interpreted as parameters of the plane approximating $\phi_{\mathrm{HWP},n}$. $N_x$ and $N_y$ are the components describing the joint effect of atmospheric and photon noise.

\begin{figure}
	\begin{center}
	\includegraphics[width=8cm]{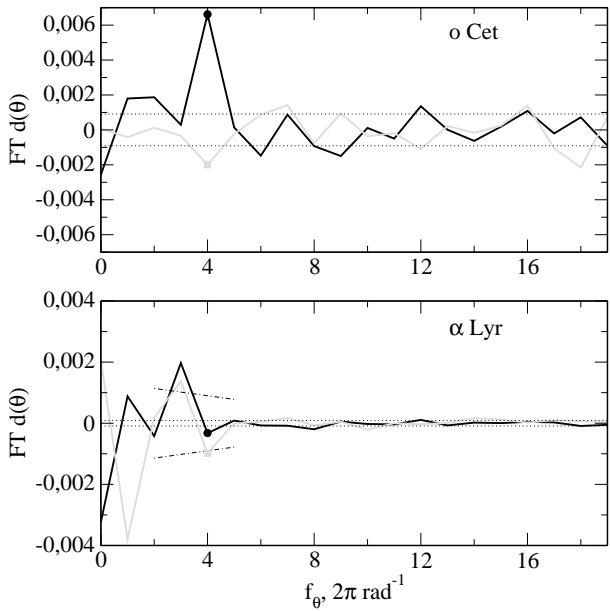}
	\end{center}
	\caption{
	Fourier spectra $\widetilde{d}(\theta)$ for $o$~Cet and $\alpha$~Lyr. Black lines are for the real part and grey are for the imaginary part. The dashed and dash-dotted lines depict level of harmonics RMS estimated from the high- and low-frequency parts, respectively.
	\label{fig:vega_mira_spec}
	}
\end{figure}

After estimation of $c(\theta)$ and $d(\theta)$ for different $\theta$ values we decompose them into a Fourier series by this parameter:
\begin{multline}
c(\theta) = s_q^{\star} \cos(4\theta) + s_u^{\star} \sin(4\theta) \\ + \sum_{n=1}^{\infty} g_q(n) \cos(n\theta) + \sum_{n=1}^{\infty} g_u(n) \cos(n\theta),
\label{eq:astromModelX}
\end{multline}
\begin{multline}
d(\theta) = t_q^{\star} \cos(4\theta) + t_u^{\star} \sin(4\theta) \\ + \sum_{n=1}^{\infty} h_q(n) \cos(n\theta) + \sum_{n=1}^{\infty} h_u(n) \cos(n\theta).
\label{eq:astromModelY}
\end{multline}
Here the effect of the HWP imperfection and random noise is decomposed in Fourier series $g_q$, $g_u$, $h_q$, and $h_u$. From the equations one can see that the behaviour of $c(\theta)$ and $d(\theta)$ is very similar to the dependence of $P(\theta)$ in the polarimetry method described earlier (Section  \ref{sec:polarimetry}, Eq. (\ref{eq:polfrac})). Therefore we analyze them by analogy with Section \ref{sec:polarimetry}. In both formulae the first two components correspond to the signal and the sums correspond to the noise input. The signal is estimated as a complex amplitude of fourth harmonics and its uncertainty is estimated by interpolation of the noise spectrum between adjacent frequencies.

In Fig. \ref{fig:vega_mira_spec}, the Fourier spectra $\widetilde{d}(\theta)$ for $o$~Cet and $\alpha$~Lyr are presented; total number of accumulated photoelectrons is $1.5\times10^{9}$ and $3.8\times10^{11}$, respectively. In the spectrum for $\alpha$~Lyr, it is easy to distinguish two regions -- of low- and high-frequency, divided by approximately the fifth harmonics. The high-frequency part results from photon noise and other uncorrelated noises, whereas the low-frequency part can be explained by the HWP imperfections (see Section \ref{subs:noise} and Appendix \ref{app:hwp}). For the low-frequency noise dominance, to estimate the fourth harmonic amplitude uncertainty one should interpolate the noise from adjacent frequencies (dash-dotted line in Fig. \ref{fig:vega_mira_spec}). For faint objects, such as $o$~Cet, it is possible to use a mean high-frequency noise amplitude for this purpose. Taking this into account makes clear that $\alpha$~Lyr does not show a distinct polaroastrometric signal whereas the $o$~Cet does. In the next subsection we analyze briefly the properties of the noise components of the spectra $\widetilde{c}(\theta)$ and $\widetilde{d}(\theta)$.

The obtained values $s_q^{\star}, s_u^{\star}, t_q^{\star}$, and $t_u^{\star}$ and their errors are converted to the equatorial coordinate system using relations (\ref{eq:unreducedRot}) from Appendix \ref{app:rotation}. The results and their discussion are given in Section \ref{sec:results}. 

\subsection{Polaroastrometric signal noise}
\label{subs:noise}

\begin{figure}
	\begin{center}
	\includegraphics[width=7cm]{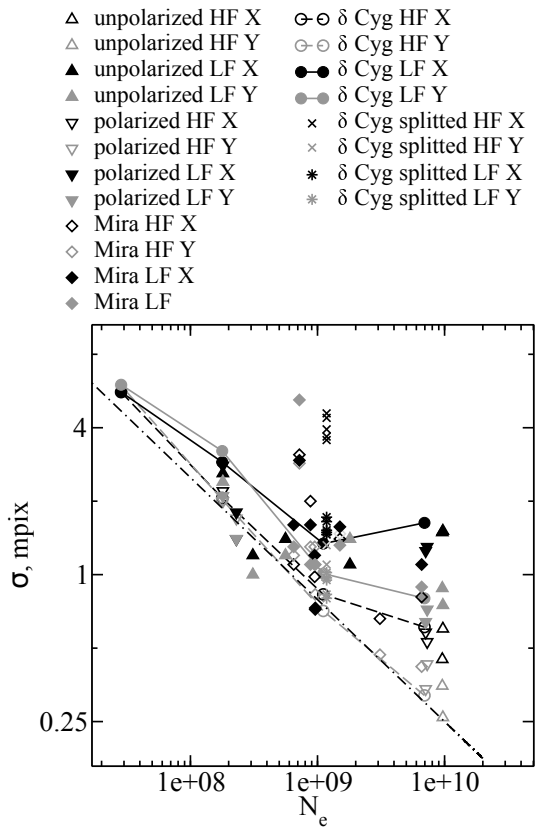}
	\end{center}
	\caption{
	Dependence of the polaroastrometric signal $\Delta$ uncertainty on the total number of accumulated photoelectrons. The black symbols are for the $X$ component, the grey ones are for the $Y$ component, empty symbols are for the high-frequency error, and filled ones for the low-frequency error (see text). The dash-dotted line is for the photon noise level of astrometry $25 \mathrm{pix}/\sqrt{N_e}$.
	\label{fig:noise_mag}
	}
\end{figure}

As a first step we identify the dependence of noise estimations from the high- and low- frequency parts of the spectra $\widetilde{c}(\theta)$ and $\widetilde{d}(\theta)$ on the object brightness. The low- and high-frequency errors are denoted as $\sigma_{\mathrm{HF}}$ and $\sigma_{\mathrm{LF}}$, respectively. For analysis we use the data for the relatively bright star $\delta$~Cyg (see Section \ref{subs:nonpol}). Additional Poisson noise is introduced into the raw data, effectively dimming the object by 2, 4, and 6 magnitudes. After this, the processing described in the previous section is applied and $\sigma_{\mathrm{HF}}$ and $\sigma_{\mathrm{LF}}$ are estimated. The results are given in Fig. \ref{fig:noise_mag} for $X$ and $Y$ coordinates separately. In this subsection it is assumed that the $X$ coordinate is measured along the CCD rows.

As one can see, $\sigma_{\mathrm{HF}}$ depends on the photoelectrons number $N_e$ as $25/\sqrt{N_e}$~pix (dash-dotted line in Fig. \ref{fig:noise_mag}), which illustrates the fact that the high-frequency part of the spectra $\widetilde{c}(\theta)$ and $\widetilde{d}(\theta)$ is produced by photon noise. It is instructive to compare this with the photon limit of astrometry noise, which is $\beta/\sqrt{N_e}$ \citep{Lindegren1978}, where $\beta$ is the full width at half maximum (FWHM) of the image. As long as light is split into two halves in our method, and then the half-difference between image centroids is taken, for the considered signal the photon limit of astrometry noise is also $\beta/\sqrt{N_e}$. Hence the high-frequency noise of the polaroastrometric signal determination corresponds to the photon limit of astrometry noise at $\beta=25$~pix or $\beta=1.7^{\prime\prime}$, which is comparable to the FWHM of a long-exposure image. 

Photon noise dominates at $N_e\lesssim10^{9}$; for our telescope and instrument this corresponds to a total integration of 500~sec from an object $V\approx6$~mag. For larger $N_e$ the low-frequency part $\sigma_{\mathrm{LF}}$ gives a larger input. The latter varies from 0.6 to 1.5~mpix and does not show a clear dependence on $N_e$.

Let us consider the correlation properties of the errors, for which the series for $\delta$~Cyg is split into five equal parts, with $\sigma_{\mathrm{HF}}$ and $\sigma_{\mathrm{LF}}$ being estimated for each separately (see Fig. \ref{fig:noise_mag}). From the figure it is evident that the high-frequency noise is uncorrelated on timescales comparable to the series duration and can be reduced by increasing the accumulation time. Meanwhile, for the low-frequency noise this is not correct. Therefore we can conclude that a sample of the low-frequency error within a single series is approximately constant. % On the other hand it changes from series to series and cannot be calibrated out.

As described in Appendix \ref{app:hwp}, the low-frequency error can be explained by the imperfection of the HWP given that the centre of the beam passing through it deviates from its centre by 1~mm. Such a decentring value is to be expected because the mechanical structure of the instrument is susceptible to flexures. This also explains why the low-frequency error is highly correlated for a single series of some object but changes after pointing to another object.

% potentially can be reduced by an alignment of the HWP rotation axis and a centre of a beam. This can be achieved by installation of HWP in an exit pupil plane or by guiding of object. It is necessary to increase the stiffness of the instrument structure in order to limit displacements of beam on the HWP to 0.1~mm in any position of instrument relative to vector of gravity. In case of effectiveness of these measures the HWP imperfection effect will be reduced to 0.1~mpix, which corresponds to 7~$\mu$as, 2~$\mu$as and 0.8~$\mu$as, for 70-cm, 2.5-m and 6-m telescope, respectively.

% Such value of decentering is quite expectable for two reasons. First, the position of an illuminated portion of the HWP depends on the position of image in the first focal plane because the HWP is installed not in the exit pupil plane of the system. Secondly, the mechanical structure of the device is not stiff enough (Sections \ref{subs:pa:extraction} and \ref{sec:discussion}).

% DO NOT DELETE THIS
% It is also remarkable that the error in $X$ coordinate generally larger than in $Y$. This is caused by the fact that the $X$--separation of the images formed by the Wollaston prism depends on $X$ component of the beam angle of incidence. Therefore any motion of the image in the focal plane (errors of tracking, vibrations, atmospheric jitter) leads to an increase of error in $X$. Apart from this, dispersion of Wollaston prism is directed along $X$ axis and an amplitude of the spectrum falls more rapidly in this direction, which also leads to increase in error.

In Fig. \ref{fig:noise_mag} $\sigma_{\mathrm{HF}}$ and $\sigma_{\mathrm{LF}}$ for some objects considered in Section \ref{sec:results} are given. As one can see polarized stars and Mira type variables exhibit error properties similar to those of unpolarized stars, except for sporadic cases.

\section{Results}
\label{sec:results}

To test the method efficiency we observed objects of several types: presumably unpolarized stars (type N), stars polarized by interstellar dust (type P), and Mira type pulsating variables (type M). The observation method and data processing were identical for all objects. Observation circumstances and processing results (polaroastrometric signal and Stokes parameters) are given in Table \ref{table:polaroastrometry}.

\subsection{Unpolarized stars}
\label{subs:nonpol}

Main sequence stars presumably have a very small polaroastrometric signal owing to their symmetry and low polarization even close to their limb. We observed three single stars (HD184960, HD221477, and HD9562), satisfying the following criteria: spectral class from A to M, luminosity class V, distance to the Sun $<66$~pc and galactic latitude $>10^{\circ}$. HD9562 exhibits a small total polarization which could be caused by interstellar dust (an effect sometimes detectable even for nearby stars \citep{Bailey2010}).

Three additional observed stars satisfy only some of the listed criteria and have peculiarities but nevertheless they can be classified as unpolarized stars. HD180161 is a BY Dra type variable of spectral class G8V and has an angular diameter of 330~$\mu$as. Brightness variations for BY Dra variables are caused by starspots occupying a large part of the stellar surface and by rotation of the star. Polarization at the limb of its disc can be crudely assumed to be the same as solar, i.e., $\approx10^{-3}$ \citep{FluriStenflo1999}. Therefore for the most favourable geometry the expected total polaroastrometric signal is 0.3~$\mu$as, which is much smaller than the error of our measurements. % DO NOT DELETE THIS Meanwhile the total polarization for this star is clearly detected in our experiment (see Table \ref{table:polaroastrometry}). However, a single measurement doesn't allow to make a conclusion about origin of polarization, it can be attributed to both the star itself and interstellar dust.

HD210027 = $\iota$~Peg is a spectroscopic binary system. The angular size of the semi-major axis of its orbit is $10.3$~mas, its period is $10.2^d$, and the spectral classes of its components are F5V and G8V \citep{Konacki2010}. One can expect a small total polarization resulting from its ellipticity but we did not detect it. Its polaroastrometric signal is likely to be negligible given the object's symmetry.

The main component of the spectroscopic binary HD186882 = $\delta$~Cyg is brighter than its secondary by $3.4$~mag; the former has a spectral class of B9.5IV and falls into the class of rapidly rotating stars. The expected angular diameter is 550~$\mu$as, and the expected degree of polarization at the limb is $\lesssim5$ per cent \citep{Bochkarev1985}. Therefore the total expected polaroastrometric signal for this star is $\lesssim25$ $\mu$as, which is undetectable with our instrument. According to \citet{Bailey2010} the total polarization for HD186882 is $(1.08\pm0.02)\times10^{-4}$.

For the six mentioned stars the expected polaroastrometric signal is much less than the errors of our measurements. This was proven in the experiment, see Table \ref{table:polaroastrometry}. This allows us to argue that our measurements do not have any bias at least down to $100~\mu$as. A negligible value of bias of this kind is expected if one takes into account the central symmetry of the optical scheme.
 
\begin{table*}
\caption{
Circumstances of the observations, the polaroastrometric signal and the dimensionless Stokes parameters for some observed objects. Column 1 lists the number in Henry Draper catalogue or Bayer name or Argelander name; Column 2 lists the universal time MM-DD HH:MM for 2014; column 3 lists the session number; column 4 lists the mean zenith distance during the series; column 5 lists the mean long-exposure FWHM during the series; column 6 lists the total number of accumulated photoelectrons divided by $10^{8}$; columns 7--10 lists the polaroastrometric signal in J2000 coordinates; and columns 11 and 12 list the dimensionless Stokes parameters in J2000 coordinates.
% processing without background because it is negligible
\label{table:polaroastrometry}}
\centering
\newcolumntype{d}[1]{D{,}{\pm}{#1}}
\tabcolsep=0.14cm
\begin{tabular}{ccccccd{3}d{3}d{3}d{3}d{3}d{3}}
\hline
HD/name & UT & S & $z,^{\circ}$ & $\beta,^{\prime\prime}$ & $N_e^{\prime}$ & \multicolumn{1}{c}{$s_q^{\star}$, $\mu$as} & \multicolumn{1}{c}{$t_q^{\star}$, $\mu$as} & \multicolumn{1}{c}{$s_u^{\star}$, $\mu$as} & \multicolumn{1}{c}{$t_u^{\star}$, $\mu$as} & \multicolumn{1}{c}{$q\times10^4$} & \multicolumn{1}{c}{$u\times10^4$} \\
1 & 2 & 3 & 4 & 5 & 6 & 7 & 8 & 9 & 10 & 11 & 12\\
\hline
\multicolumn{12}{c}{\rule{0pt}{2.5ex} unpolarized stars} \\
%\hline
186882 & 08-03 19:53 & 1 & 11.7 & 2.2 & $69$ & +20,30 & +10,50 & +40,30 &  +0,50 & 0.8,0.5 & 0.8,0.5 \\ % delta Cygni -- unpol
184960 & 08-03 20:12 & 1 & 4.4 & 2.2 & $5.6$ & +130,90 & -70,100 &  -0,90 & -90,100 & -1.0,0.8 & -0.5,0.8 \\  % HIP96258 -- unpol
180161 & 08-04 18:42 & 1 & 9.3 & 1.8 & $3.1$ & +50,70 & +70,80 & +120,70 & -10,80 & -4.7,0.9 & 3.4,0.9 \\  % HIP94346 -- unpol
221477 & 08-04 22:45 & 1 & 24.8 & 2.7 & $1.8$ & -40,180 &  +0,180 & +100,180 & +90,180 &  0.3,1.8 & -3.9,1.8 \\ % HIP116174 -- unpol
210027 & 10-02 19:51 & 5 & 32.2 & 2.0 & $97$ & +110,80 & +10,90 & -50,80 & -70,90 &   0.3,0.4 & -1.3,0.4 \\ %HIP109176 top of frame -- unpol
210027 & 10-02 20:11 & 5 & 33.6 & 2.1 & $96$ &  +90,80 & +20,90 & +10,80 & -100,90 &  -0.6,0.3 & 0.3,0.3 \\ %HIP109176 bottom of frame -- unpol
9562   & 10-02 21:16 & 5 & 63.9 & 3.9 & $18$ & +70,100 & +10,100 & +210,100 & -40,100 & 3.5,0.5 & -1.4,0.5 \\ %HIP7276 -- unpol
%\hline
\multicolumn{12}{c}{\rule{0pt}{2.5ex} stars polarized by interstellar dust} \\
%\hline
204827 & 08-04 20:35 & 1 & 12.5 & 2.0 & $2.3$ & +90,120 & +110,120 & +20,120 & +50,120& -258.2,5.3 & 483.0,3.2 \\ % HIP106059 pol. standard
204827 & 09-19 19:00 & 4 & 3.2 & 2.1 & $1.8$ & +70,180 & -20,160 & -140,180 & +210,160 & -282.7,5.6 & 471.3,3.7 \\ % HIP106059 pol. standard
7927   & 09-18 16:45 & 3 & 46.9 & 2.0 & $73$ & +10,70 & -60,80 & -30,70 & -150,80 & -333.1,0.7 & -31.1,4.3 \\ % phi Cas pol. standard
7927   & 09-19 17:44 & 4 & 40.1 & 2.2 & $71$ & -100,70 & +200,60 & -100,70 & +120,60 & -328.4,1.2 & -35.6,4.0 \\ % phi Cas pol. standard
%\hline
\multicolumn{12}{c}{\rule{0pt}{2.5ex} Mira type variables} \\
%\hline
$\chi$~Cyg & 08-03 21:24 & 1 & 24.5 & 2.6 & 9.5 & +230,90 & +210,80 & -50,90 & -240,80 & -22.5,0.8 & -28.9,0.7 \\ % chi cyg P.A. of zen. direction 18.0, chicygFinal.m JD=2456873.39167, phase=0.26
$\chi$~Cyg & 08-04 19:16 & 1 & 25.9 & 2.3 & 9.6 & +220,70 & +220,50 & -0,70 & -320,50 & -23.9,0.6 & -31.5,0.6 \\ % chi cyg P.A. of zen. direction -23.2, chicygFinal.m JD=2456874.30278, phase=0.26
$\chi$~Cyg & 09-18 18:05 & 3 & 23.6 & 2.0 & 6.5 & +500,100 & +430,110 & -60,100 & +50,110 & -23.1,1.1 & -40.6,1.0 \\ % chi cyg P.A. of zen. direction 12.4, chicygFinal.m, JD=2456919.25347, phase=0.37
R~Tri  & 09-18 21:45 & 3 & 30.6 & 1.8 & 31 & -50,40 & -60, 40 & -30, 40 & -180, 40 & 20.5,1.0 & -74.1,0.5\\ % R Tri, rtriFinal.m, JD=2456919.40694, phase=0.91
R~Tri  & 09-20 21:37 & 4 & 31.1 & 2.4 & 66 & +20,70 & +80, 70 & +30, 70 & +30, 70 & 21.0,1.1 & -75.7,0.7 \\ % R Tri, rtriFinal.m, JD=2456921.40139, phase=0.92
$o$~Cet  & 09-20 23:01 & 4 & 60.0 & 3.2 & 15 & -150,110 & -470,100 & +890,110 & -790,100 & -1.5,0.7 & -9.5,0.7 \\ % Mira, miraFinal.m, JD=2456921.45903, phase=0.54
$o$~Cet  & 10-03 21:55 & 5 & 60.2 & 2.9 & 8.8 & +10,110 & -430,120 & +850,110 & -750,120 & 0.9,0.8 & -15.0,0.8 \\ % Mira, miraFinal.m, JD=2456934.39931, phase=0.58
$o$~Cet  & 10-03 22:37 & $5^{\prime}$ & 58.8 & 3.1 & 7.2 & -130,350 & -440,240 & +1050,350 & -740,240 & 1.4,1.4 & -17.0,1.4 \\ \hline
\end{tabular}
\end{table*}

\subsection{Stars polarized by interstellar dust}
\label{subs:atmdisp}

An undetectable polaroastrometric signal is expected for stars analogous to type N but polarized by interstellar dust (type P), because the latter changes the polarization of each point of a star image identically. However, a small signal can be generated for observations made through the Earth's atmosphere owing to the joint effect of dependence of the degree of polarization on the wavelength and atmospheric dispersion.

Indeed, for some polarization standards variation of the degree of polarization across the $V$ band reaches $0.2\div0.3$ per cent in absolute measure \citep{Schmidt1992}. Hence the polarized flux has an effective wavelength differing from the effective wavelength of the total flux by $\Delta\lambda$. For computing the effective wavelength we used the stellar spectra of \citet{Gunn1983}, the sensitivity curve of the detector, the transmission curve of the filter, and the dependence of degree of the polarization on the wavelength for the considered stars from \citet{Schmidt1992}. Differences in the effective wavelengths are $\Delta\lambda=-1.027$~nm and $\Delta\lambda=-0.480$~nm for HD204827 and HD7927, respectively. These stars also display a dependence of the angle of polarization on the wavelength \citep{Schmidt1992} but this can be neglected since its maximum amplitude over the $V$ band is $0.8^{\circ}$.

Atmospheric dispersion displaces the polaricentres and this displacement is the same for polaricentres $q$ and $u$, as long as $\Delta\lambda$ is the same. The total displacement is $\sqrt{s_q^2+t_q^2} = \sqrt{s_u^2+t_u^2} = \epsilon\cdot\Delta\lambda\cdot \mathrm{tan}z$, where $\epsilon$ is the angular atmospheric dispersion ($\epsilon=2.12\times10^{-8}$~rad/nm for place and time of observations) and $z$ is zenith distance. The amplitude of polaroastrometric signal can be estimated according to its definition. At $z=45^{\circ}$ for HD204827 and HD7927 the amplitudes are 254 and 69~$\mu$as, respectively. In our experiment such a small signal cannot be reliably measured.

\subsection{Mira type variables}
\label{subs:chicyg}

Mira type variables are pulsating stars at late stages of evolution. These objects are expected to demonstrate detectable deviations of their polaricentres from the photocentre for two reasons. Firstly, a significant fraction of the visible light radiation is being scattered by the dust envelope in the close vicinity of a star \citep{Norris2012}, which leads to polarization of radiation. Secondly, interferometric observations show that some Mira variables display significant asymmetry \citep{Thompson2002,Ragland2006forMNRAS}.

We considered three Mira variables, suitable for observations in the northern hemisphere from August to October of 2014: $\chi$~Cyg, R~Tri, and $o$~Cet. The measured total polarization and polaroastrometric signal are given in Table \ref{table:polaroastrometry}. For R~Tri we did not detect a significant signal; however, the total polarization for this star is largest in comparison with the other observed Mirae. This can be explained by the axial symmetry of this object detected previously by \citet{Thompson2002} using the Palomar Testbed Interferometer in near infrared and confirmed by \citet{Ragland2006forMNRAS} using the Infrared Optical Telescope Array in the same region of the spectrum. 

$\chi$~Cyg and $o$~Cet show significant deviation of values $s_q^{\star}, s_u^{\star}, t_q^{\star}$, and $t_u^{\star}$ from zero, indicating the asymmetry of image, the same conclusion had been reached by \citet{Ragland2006forMNRAS} for $\chi$~Cyg and by \citet{Karovska1997,Chandler2007} for $o$~Cet (in the visible and mid-infrared). Moreover, for these objects the polaricentres $q$ and $u$ clearly do not coincide: $s_q^{\star}/q \ne s_u^{\star}/u$ and $t_q^{\star}/q \ne t_u^{\star}/u$. Therefore the polaroastrometric signal cannot be explained by a single polarized source.

For $o$~Cet the situation is complicated by the presence of the nearby component Mira~B, an accreting white dwarf \citep{Sokoloski2010}, which in principle can be polarized. On 2014 October 3, 21:55 UT the components of the $u$--polaricentre were measured to be $s_u = -570\pm80$~mas and $t_u = 500\pm80$~mas; in the same reference frame the position of Mira~B was $s_{c} = -70$~mas and $t_{c} = 480$~mas \citep{Prieur2002}. Therefore the $u$--polaricentre was shifted in roughly the same direction as Mira~B, which can be interpreted as the effect of the polarized emission of Mira~B seen against the background of the ``random'' polaroastrometric signal generated by the Mira~A atmosphere. A more definitive conclusion on polarimetric properties of Mira~B in the visible requires more observations conducted in different phases of the pulsations.

% $\chi$~Cyg and $o$~Cet show significant deviation of values $s_q^{\star}, s_u^{\star}, t_q^{\star}, t_u^{\star}$ from zero, for R~Tri it is not so; meanwhile the total polarization for R~Tri is the largest. These results conform qualitatively to measurements of \citet{Ragland2006forMNRAS}, who estimated asymmetry of stars images by measuring closure phases formed by three baselines of the IOTA interferometer in NIR. For $\chi$~Cyg significant deviation of the closure phase from $0^{\circ}/180^{\circ}$ was detected what points out to the asymmetry of image, R~Tri was found to be symmetrical. \citet{Thompson2002} also showed using the PTI interferometer that R~Tri has axial symmetry. The axial symmetry can explain the situation when total polarization is large while polaroastrometric signal is undetectable. $o$~Cet earlier was found to be significantly asymmetrical in visible light \citep{Karovska1997}, as well as in MIR \citep{Chandler2007}. 

% DO NOT DELETE THIS
%The angular diameters of the considered stars are $\approx22$~mas for $\chi$~Cyg \citep{Ragland2006forMNRAS}, $\approx22~$mas for $o$~Cet \citep{Woodruff2004forMNRAS} and 4.6~mas for R~Tri \citep{Thompson2002}.

To test the methodology we repeated observations of $\chi$~Cyg and $o$~Cet in nearby epochs; corresponding measurements agree well (see Table \ref{table:polaroastrometry}). In case of $o$~Cet, between the second and third measurements the instrument was rotated counterclockwise with respect to the telescope by $57.5^{\circ}$. As one can see $s_q^{\star}, s_u^{\star}, t_q^{\star}$, and $t_u^{\star}$, transformed to the equatorial coordinate system, remained the same after rotation. For a larger difference in epochs -- $55.0^d$ (between second and third measurements of $\chi$~Cyg), the polaroastrometric signal changes significantly. It is worth noting that the brightness of $\chi$~Cyg for the same period decreased from $V=7.5$~mag to $V=8.5$~mag (AAVSO data).

\section{Conclusions}
\label{sec:conclusion}

The goals of the present work are to develop a methodology and to study the precision limits of polaroastrometry. Analysis has been performed on the basis of observational data obtained with the help of the Multimode Fast Camera prototype. By analysing the dependence of the half-difference of the photocentres of the images split by a Wollaston prism on the position angle $\theta$ of the HWP we obtained estimations of the polaroastrometric signal, i.e., displacements of the centroids of the polarized flux $Q$ or $U$ from the photocentre of the total flux, multiplied by the corresponding dimensionless Stokes parameter, $q$ or $u$.

We performed an analysis of the polaroastrometric signal error dependence on circumstances of an experiment. It was found that for a total number of accumulated photoelectrons $N_e < 10^9$ the photon noise dominates and can be approximated crudely as $1.7^{\prime\prime}/\sqrt{N_e}$. At larger $N_e$ the input of imperfections of the given HWP starts to prevail, having an approximately constant level of $50$--$100~\mu$as. This effect can be reduced by an order of magnitude if the HWP rotation and beam centre are aligned with a precision of 0.1~mm (see Appendix \ref{app:hwp}). %In Fig. \ref{fig:noiseV} we give a rough estimation of the photon noise and HWP imperfections noise for larger telescopes.

Observations of unpolarized main sequence stars, which presumably do not have any detectable polaroastrometric signal, showed that the measurements are unbiased at least at the level $\approx100~\mu$as. Main sequence stars polarized by interstellar dust also do not display a detectable signal demonstrating the absence of cross-talk between a strong polarization and the polaroastrometric signal.

For method verification purposes we selected three Mira type variables $\chi$~Cyg, $o$~Cet, and R~Tri because they are expected to have detectable signal. For $\chi$~Cyg the measured polaroastrometric signal is $s_q^{\star}=+220\pm70$ $\mu$as, $t_q^{\star}=+220\pm50$ $\mu$as, $s_u^{\star}=0\pm70$ $\mu$as, and $t_u^{\star}=-320\pm50$ $\mu$as in the J2000 coordinates for 2014 August 4, 19:16 UT. For $o$~Cet these values are $s_q^{\star}=+10\pm110$ $\mu$as, $t_q^{\star}=-430\pm120$ $\mu$as, $s_u^{\star}=+850\pm110$ $\mu$as, and $t_u^{\star}=-750\pm120$ $\mu$as for 2014 October 3, 21:55 UT. A significant polaroastrometric signal gives evidence of asymmetry of the polarized visible flux. Asymmetry was detected for $\chi$~Cyg and $o$~Cet earlier in the infrared by \citet{Ragland2006forMNRAS} and \citet{Chandler2007}, and for the latter in the visible also by \citet{Karovska1997}. For R~Tri the signal was not found.

As a test of the repeatability of measurements we observed these objects at different, but close epochs and at different orientations of the instrument relative to the telescope. We did not detect a significant difference in measurements. Also as a by-product we obtained the polarimetry of the mentioned objects with a precision of $(1\div5)\times10^{-4}$.  

Mira type variables are suitable as test objects because apparently they have the greatest polaroastrometric signal among all celestial objects. %For the relatively small and unfortunately located 70-cm telescope these variable stars are only objects suitable for measurements. 
Though the obtained data demonstrate the effectiveness of the method, their astrophysical interpretation is obstructed by the complexity of these objects.

Possible applications of polaroastrometry using a large telescope to polarized objects described by a simple geometrical model, e.g., some types of circumstellar environments, active galactic nuclei, including Sgr A$*$, and gravitational microlensing are very promising.

%There are some very promising possible applications of polaroastrometry using a large telescope to .

% DO NOT DELETE THIS Let us consider, as an example, object Sgr A$*$ in centre of Milky Way which assumed to be a supermassive black hole. Observations in NIR showed that Sgr A$*$ is susceptible to irregular flares with a characteristic timescale of 10-20 minutes. During the flare the brightness of object increases from 2~mJy to $\approx5$~mJy, and the object becomes polarized at the level of $\approx10\%$ \citep{Eckart2006}. The main hypothesis explaining these flares is rotation of hot spots emitting synchrotron radiation in an accretion disk very close to the black hole. The expected level of polaroastrometric photon noise for this object at 8-m class telescope, integration time 3~min, is 100~$\mu$as and 10~$\mu$as with and without adaptive optics, respectively. The latter value is comparable to expected angular size of an event horizon for this black hole, which is 10~$\mu$as. Hence using polaroastrometry in principle it is possible to put additional geometrical constraints on models of these hot spots.

\subsection*{Acknowledgements}

This work was supported by M.V.Lomonosov Moscow State University Program of Development and RFBR Grant No. 14-04-31185. The author wishes to thank N.~Shatsky, V.~Kornilov, and A.~Tokovinin for discussions on methodology, and A.~Magnitskiy and D.~Cheryasov for help with technical issues during observations. Comments by an anonymous referee helped to improve the presentation and extend the analysis. Use was made of the SIMBAD and AAVSO databases for preparation of observations and data processing.
\bibliography{AZT2astrom} 
\bibliographystyle{mn2e}

\appendix

\section{Influence of half-wave plate imperfection}
\label{app:hwp}

\begin{figure*}
	\centering
	\includegraphics[width=16cm]{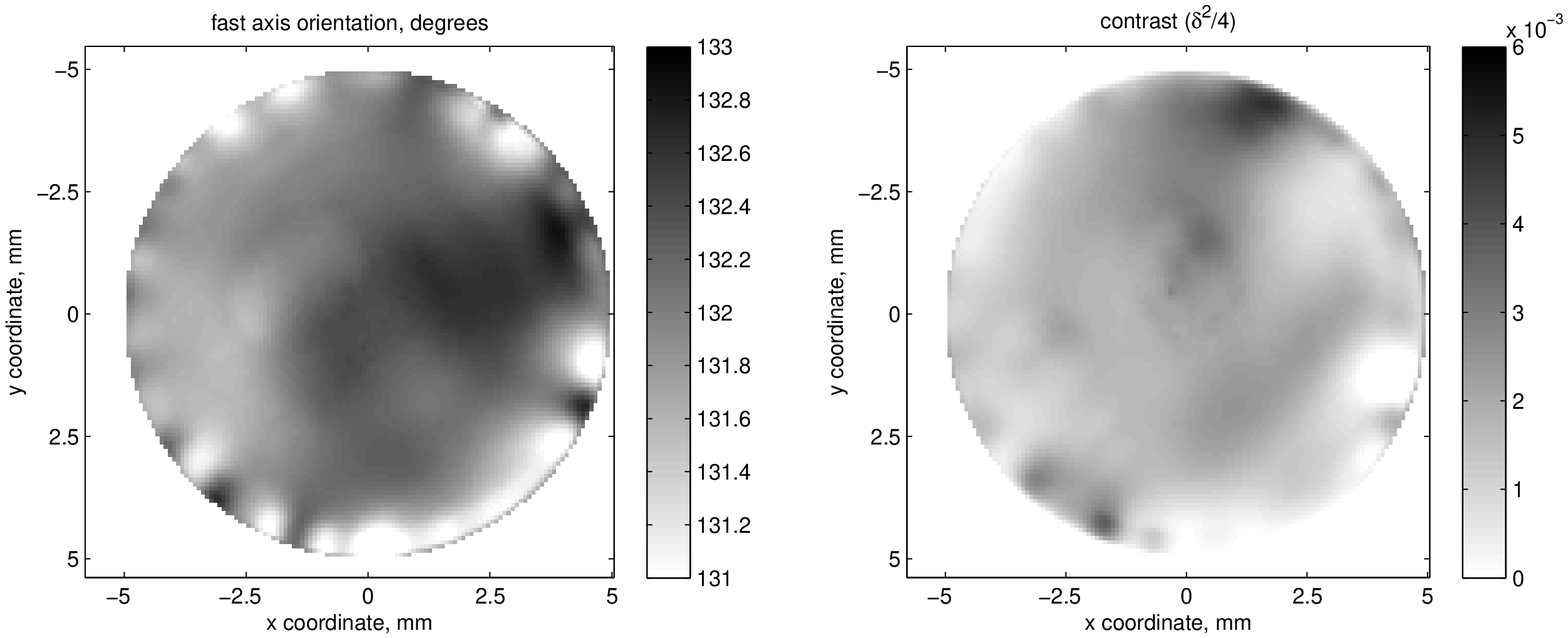}
	\caption{
	Fluctuations of the basic parameters of the HWP over its area. The left panel is for the fast axis orientation; the right panel is for the contrast at crossed linear polarizers $\delta^{\prime2}/4$, where $\delta^{\prime}$ is the deviation of the retardance from $\pi$.
	\label{fig:axis_contrast}
	}
\end{figure*}

Let us suppose that the HWP imperfection can be modeled as fluctuations of a phase retardance $\delta$ and fast axis orientation $\theta$ over its area. To derive these fluctuations for the given HWP we illuminated it through a linear polarizer, then passed the resulting light through another linear polarizer, which we could rotate. With the help of a relay lens the HWP surface was imaged on the detector. For each point on the HWP, values $\delta$ and $\theta$ were extracted from the dependence of the flux on the second linear polarizer orientation. The results are given in Fig. \ref{fig:axis_contrast}.

As one can see from figure, fluctuations of $\theta$ reach $2^{\circ}$, and become more pronounced closer to the border of the plate. For $\delta$, the amplitude of the fluctuation is of the order of $3\div5^{\circ}$ in a central part of the HWP. On the basis of these measurements and with the help of the model developed by \citet{Safonov2013} we estimated $\mathcal{R}_{\mathrm{HWP}}(\theta,\boldsymbol{f})$ (for the definition see Section \ref{subs:pa:extraction}) for different beam parameters. Differential aberrations produced by the HWP are purely phase aberrations, and the same is true for $\mathcal{R}_{\mathrm{HWP}}(\theta,\boldsymbol{f})$. The RMS of the $\mathcal{R}_{\mathrm{HWP}}(\theta,\boldsymbol{f})$ phase in area $|f|<f_c$ is 0.02 rad, in $|f|<0.2 f_c$ it is 0.002 rad.

\begin{figure}
	\centering
	\includegraphics[width=8cm]{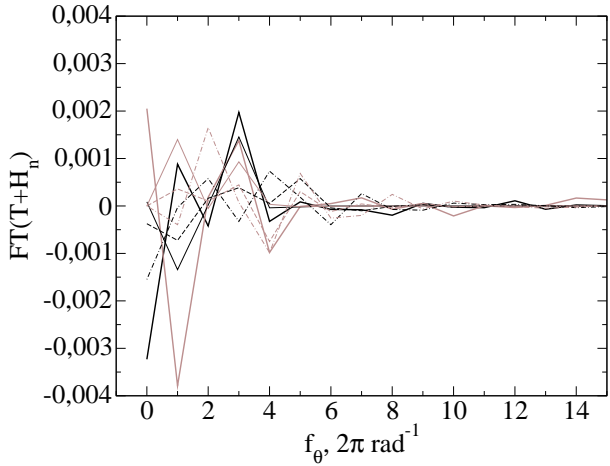}
	\caption{
	Spectrum of $H_n(\theta)$, black and grey lines show the real and imaginary parts, respectively. Bold lines are for measurements of $\alpha$~Lyr, and thin lines are for what is expected when the HWP is used. Decentring values 0, 1, and 2~mm are shown by solid, dashed, and dash-dotted lines, respectively.}
	\label{fig:HWPspec}
\end{figure}

By approximating the phase of $\mathcal{R}_{\mathrm{HWP}}(\theta,\boldsymbol{f})$ with a plane we evaluated functions $G(\theta)$ and $H(\theta)$ (see Section \ref{subs:pa:extraction}) and then computed their Fourier spectra. These are given in Fig. \ref{fig:HWPspec} for a beam diameter 2~mm and different decentring parameter in comparison with the analogous spectrum measured in the polaroastrometric experiment for $\alpha$~Lyr. It is evident that the observed behaviour of the spectra can be explained by the described HWP imperfection at decentring $\approx1$~mm. % Such value of decentering is quite expectable for two reasons. First, the position of an illuminated portion of the HWP depends on the position of image in the first focal plane because the HWP is installed not in the exit pupil plane of the system. Secondly, the mechanical structure of the device is not stiff enough (Sections \ref{subs:pa:extraction} and \ref{sec:discussion}).

\begin{figure}
	\centering
	\includegraphics[width=8cm]{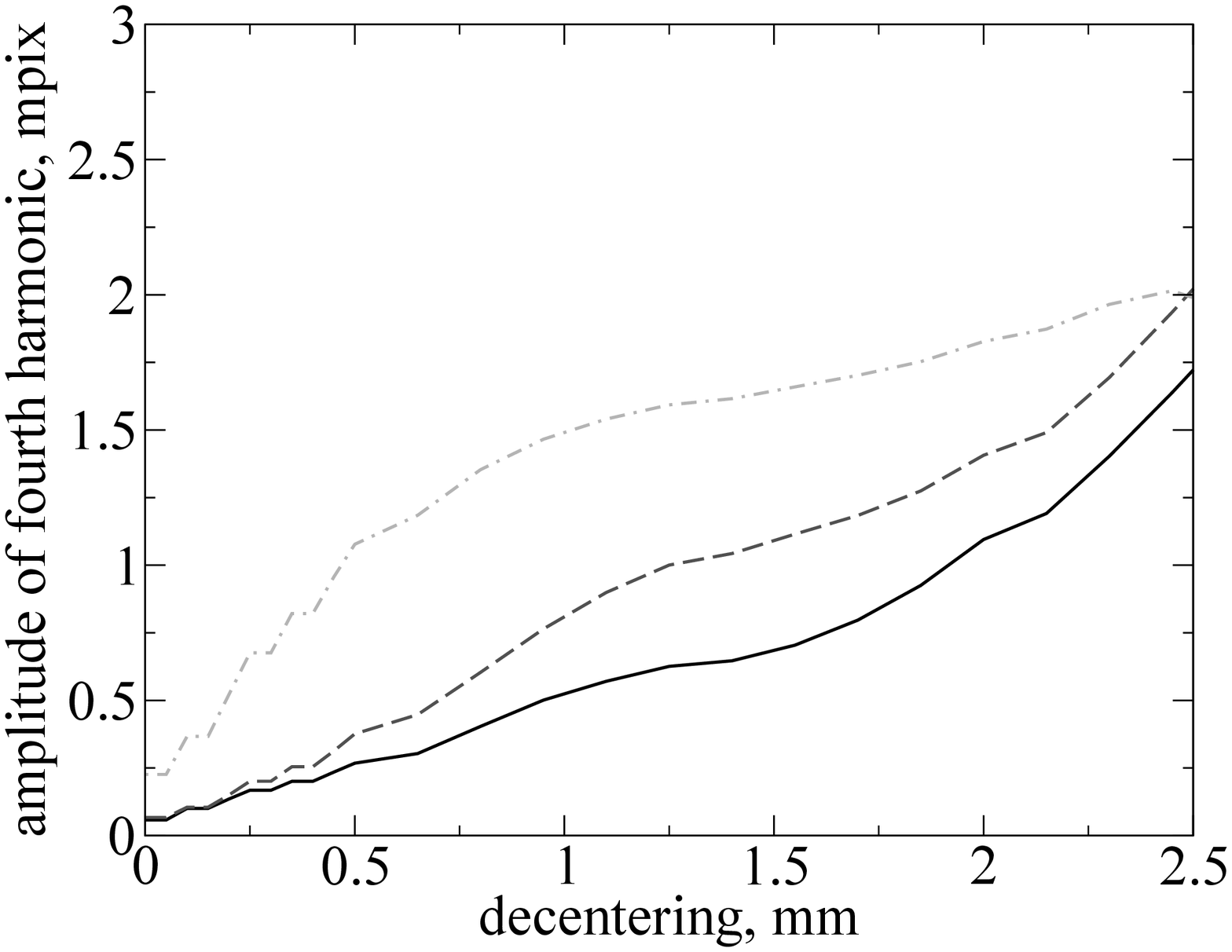}
	\caption{
	Dependence of the fourth harmonic amplitude of spectrum $H_n(\theta)$ on the decentring parameter for three values of beam diameter on the HWP: 2 (black solid line), 3 (dark grey dashed line), and 4~mm (light grey dash-dotted line).}
	\label{fig:HWPeffect}
\end{figure}

The useful signal corresponds to a complex amplitude of fourth harmonics of the Fourier spectra $\widetilde{c}(\theta)$ and $\widetilde{d}(\theta)$. The amplitude of the noise of this harmonic is given in Fig. \ref{fig:HWPeffect} as a function of beam diameter and decentring. As one can see, reducing the noise to the level of 0.1~mpix requires aligning the beam and the centre of the HWP rotation with an accuracy of 0.1~mm, it is also worthwhile to use beams of smaller diameter ($<2$~mm).

To derive the HWP imperfection influence on the polarimetry let us take its Mueller matrix in the following form (where we have kept only small first-order values and integrated components over the beam):
\begin{multline}
\boldsymbol{\mathrm{M}}_{\mathrm{HWP}} = 
\left(
\begin{tabular}{cc}
1 & 0 \\
0 & $\cos\bigl[4(\theta+\kappa(\theta))\bigr]$ \\
0 & $\sin\bigl[4(\theta+\kappa(\theta))\bigr]$ \\
0 & $-\delta(\theta)\sin\bigl[2(\theta+\kappa(\theta))\bigr]$ \\
\end{tabular}
\right. \\ 
\left.
\begin{tabular}{cc}
0 & 0 \\
$\sin\bigl[4(\theta+\kappa(\theta))\bigr]$ &  $\delta(\theta)\sin\bigl[2(\theta+\kappa(\theta))\bigr]$ \\
$-\cos\bigl[4(\theta+\kappa(\theta))\bigr]$ & $-\delta(\theta)\cos\bigl[2(\theta+\kappa(\theta))\bigr]$ \\
$\delta(\theta)\cos\bigl[2(\theta+\kappa(\theta))\bigr]$ & 1 \\
\end{tabular}
\right)
\end{multline}

In accordance with the specification, the Wollaston prism can be considered ideal in our experiment. The first rows of the Mueller matrix for left and right beams of the prism are
\begin{equation}
\boldsymbol{\mathrm{M}}_{\mathrm{WL}} = \begin{pmatrix}
1 & 1 & 0 & 0 \\
\end{pmatrix},
\end{equation}
\begin{equation}
\boldsymbol{\mathrm{M}}_{\mathrm{WR}} = \begin{pmatrix}
1 & -1 & 0 & 0 \\
\end{pmatrix}
\end{equation}
(we are interested in first rows only because we can measure the intensity only).


The total Stokes vectors of images for the left and right beams can be expressed through the matrices
\begin{equation}
\boldsymbol{S}_{\mathrm{out,L}} = \boldsymbol{\mathrm{M}}_{\mathrm{WL}} \boldsymbol{\mathrm{M}}_{\mathrm{HWP}} \boldsymbol{S}_{\mathrm{in}},
\end{equation}
\begin{equation}
\boldsymbol{S}_{\mathrm{out,R}} = \boldsymbol{\mathrm{M}}_{\mathrm{WR}} \boldsymbol{\mathrm{M}}_{\mathrm{HWP}} \boldsymbol{S}_{\mathrm{in}},
\end{equation}
where $\boldsymbol{S}_{\mathrm{in}} = (I_{\mathrm{in}}, Q_{\mathrm{in}}, U_{\mathrm{in}}, V_{\mathrm{in}})$ is the Stokes vector of incoming radiation. Fluxes for the left and right beams are
\begin{multline} 
I_{\mathrm{out,L}} = I_{\mathrm{in}} + Q_{\mathrm{in}} \cos(4(\theta_k+\kappa(\theta_k))) + U_{\mathrm{in}} \sin(4(\theta_k+\kappa(\theta_k)))  \\ + V_{\mathrm{in}} \delta(\theta_k) \sin(2(\theta_k+\kappa(\theta_k))),
\end{multline} 
\begin{multline} 
I_{\mathrm{out,R}} = I_{\mathrm{in}} - Q_{\mathrm{in}} \cos(4(\theta_k+\kappa(\theta_k))) - U_{\mathrm{in}} \sin(4(\theta_k+\kappa(\theta_k))) \\ - V_{\mathrm{in}} \delta(\theta_k) \sin(2(\theta_k+\kappa(\theta_k))).
\end{multline} 

\section{Polaroastrometric signal transformation to rotated reference system}
\label{app:rotation}

% DO NOT DELETE THIS
%Stokes parameters $q,u$ are transformed in this way:
%\begin{align}\begin{split}
%q^{\prime} & =    q \cos2\psi + u \sin2\psi, \\
%u^{\prime} & =  - q \sin2\psi + u \cos2\psi.
%\end{split}\end{align}

%Components $x,y$ of some vector are transformed this way:
%\begin{align}\begin{split}
%x^{\prime} & =   x \cos\psi + y \sin\psi, \\
%y^{\prime} & = - x \sin\psi + y \cos\psi.
%\end{split}\end{align}

For practical purposes we frequently need to transform the polaroastrometric signal to a new reference system rotated relative to the initial system by some angle $\psi$, measured counterclockwise. Taking into account the transformation of vectors and the Stokes parameters in the rotated reference system, we derived these transformations for the components of the polaroastrometric signal as follows:
\begin{align}\begin{split}
s^{\star\prime}_q = & (s^{\star}_q\cos2\psi+s^{\star}_u\sin2\psi)\cos\psi  \\
                    & + (t^{\star}_q\cos2\psi+t^{\star}_u\sin2\psi)\sin\psi, \\
t^{\star\prime}_q = & -(s^{\star}_q\cos2\psi+s^{\star}_u\sin2\psi)\sin\psi \\ 
                    & + (t^{\star}_q\cos2\psi+t^{\star}_u\sin2\psi)\cos\psi, \\
s^{\star\prime}_u = & (-s^{\star}_q\sin2\psi+s^{\star}_u\cos2\psi)\cos\psi \\ 
                    & + (-t^{\star}_q\sin2\psi+t^{\star}_u\cos2\psi)\sin\psi, \\
t^{\star\prime}_u = & -(-s^{\star}_q\sin2\psi+s^{\star}_u\cos2\psi)\sin\psi \\ 
                    & + (-t^{\star}_q\sin2\psi+t^{\star}_u\cos2\psi)\cos\psi,
\label{eq:unreducedRot}
\end{split}\end{align}
where values in new reference system are noted with primes.

For polaricentres we get
\begin{align}\begin{split}
s_q^{\prime} & = (a s_q + b s_u)\cos\psi + (a t_q + b t_u)\sin\psi, \\
t_q^{\prime} & = -(a s_q + b s_u)\sin\psi + (a t_q + b t_u)\cos\psi, \\
s_u^{\prime} & = (c s_q + d s_u)\cos\psi + (c t_q + d t_u)\sin\psi, \\
t_u^{\prime} & = -(c s_q + d s_u)\sin\psi + (c t_q + d t_u)\cos\psi,
\end{split}\end{align}
%\langru{где}
where
\begin{align}\begin{split}
a = \frac{q\cos2\psi}{q\cos2\psi+u\sin2\psi}, \\
b = \frac{u\sin2\psi}{q\cos2\psi+u\sin2\psi}, \\ 
c = \frac{-q\sin2\psi}{-q\cos2\psi+u\sin2\psi}, \\
d = \frac{-u\cos2\psi}{-q\cos2\psi+u\sin2\psi}.
\end{split}\end{align}

As one can see, sometimes it is impossible to transform the polaricentres because the denominators in the last four expressions may be close to zero. In such cases one should operate with the polaroastrometric signal itself.

It is worth noting that when the polaricentres $q$ and $u$ coincide, transformation of their coordinates is a simple rotation:
\begin{align}\begin{split}
s^{\prime} = s \cos\psi + t \sin\psi, \\
t^{\prime} =  - s \sin\psi + t \cos\psi.
\end{split}\end{align}

\label{lastpage}
\end{document}